\documentclass{article}
\usepackage[preprint]{spconf}
\copyrightnotice{978-1-6654-7189-3/22/\$31.00~\copyright2023 IEEE}

\usepackage{amsmath,amsfonts}
\usepackage{algorithm}
\usepackage{array}
\usepackage[caption=false,font=normalsize,labelfont=sf,textfont=sf]{subfig}
\usepackage{textcomp}
\usepackage{stfloats}
\usepackage{url}
\usepackage{verbatim}
\usepackage{graphicx}
\usepackage{cite}

\usepackage{amsthm,amsopn}
\usepackage{bm}
\usepackage{hyperref}
\usepackage{url}
\usepackage[capitalise]{cleveref}
\usepackage{graphicx,caption,lipsum}
\usepackage{booktabs,multirow}
\usepackage{placeins}
\usepackage{algpseudocode,algorithm}
\usepackage{color}
\usepackage[table,xcdraw]{xcolor}
\definecolor{Gray}{gray}{0.9}
\definecolor{brightturquoise}{rgb}{0.85, 1, 1}
\usepackage{bbm}
\usepackage{enumerate}   
\usepackage{adjustbox}
\usepackage[normalem]{ulem}
\usepackage{fixltx2e}
\usepackage{wrapfig}
\usepackage{threeparttable}
\usepackage{tikz}
\usepackage{booktabs}

\usepackage{seqsplit}

\title{AV-data2vec: Self-supervised Learning of Audio-Visual Speech Representations with Contextualized Target Representations}
\name{%
  \begin{tabular}{c}
    Jiachen Lian$^1$, Alexei Baevski$^{2 \star}$, Wei-Ning Hsu$^{3 \star}$, Michael Auli$^{3 \star} \thanks{${\star}$ Equal Advising. Work done at Meta AI}$
  \end{tabular}%
}

\address{$^{1}$ UC Berkeley $^{2}$ Character.AI $^{3}$ FAIR, Meta
}

\begin{document}

\newcommand{\ma}[1]{{\color{orange}#1}}
\newcommand{\wh}[1]{{\color{magenta}#1}}
\newcommand{\jl}[1]{{\color{brown}#1}}

\newcommand{\nam}{AV-data2vec}

\newcommand{\overbar}[1]{\mkern 1.5mu\overline{\mkern-4.5mu#1\mkern-4.5mu}\mkern 1.5mu}

\ninept
\maketitle
\begin{abstract}
Self-supervision has shown great potential for audio-visual speech recognition by vastly reducing the amount of labeled data required to build good systems.
However, existing methods are either not entirely end-to-end or do not train joint representations of both modalities.
In this paper, we introduce \nam{} which addresses these challenges and builds audio-visual representations based on predicting contextualized representations which has been successful in the uni-modal case.
The model uses a shared transformer encoder for both audio and video and can combine both modalities to improve speech recognition.
Results on LRS3 show that AV-data2vec consistently outperforms existing methods under all settings with the same amount of data and model size. 
% Human speech perception is multi-modal. 
% While the deep neural networks are designed to simulate the human brain mechanism for perceiving the surrounding world, current state-of-art speech recognition work focuses entirely on pure audio signals. 
% Large-scale availability of unlabeled pairwise audio-video data in the wild provides the possibility to explore the unsupervised audio-visual pretraining at scale. 
% In this work, we introduce AV-data2vec (Audio-Visual data2vec), a self-supervised audio-visual framework to jointly learn speech representations from contextualized targets via multi-modal training. 
% AV-data2vec employs a shared transformer encoder for both audio and video modalities, which is closer to human speech perception mechanism where early fusion of multi-modal signals is evidenced. 
% AV-data2vec is a universal self-supervised speech recognition framework that unifies ASR, VSR and AVSR. 
% Results on LRS3 show that AV-data2vec consistently outperform existing methods under most settings. 
\end{abstract}
\section{Introduction}

Both human speech production and perception are multimodal, producing acoustic and visual artifacts \cite{diehl2004speech_perception,macneilage2010origin}. 
% % Human speech production results in both acoustic and visual outputs which encode linguistic information~\citep{macneilage2010origin} and the human speech perception system is also multimodal in nature~\citep{diehl2004speech_perception}. 
% The human brain also processes speech as a combination of auditory, visual, and even tactile stimuli~\cite{mcgurk1976hearing, calvert1997silent,musacchia2006seeing,fowler1991listening}. 
% This is in stark contrast to state-of-the-art speech recognition research which focuses mostly on the audio signal to perform tasks such as automatic speech recognition (ASR; \cite{gulati2020conformer,baevski2020wav2vec2,hsu2021hubert,chen2022wavlm,baevski2022data2vec,radford2022whisper}).
Learning \textit{audio-visual speech representations} helps to improve the robustness and accuracy of speech recognition in both noisy and clean settings~\cite{AV-HuBERT,shi2022robust_avhubert}. 
% This joint processing is typically referred to as audio-visual speech recognition (AVSR).
% However, processing only the visual signal in visual speech recognition (VSR) also has important applications such as supporting people which lost their ability to speak due to various reasons~\cite{seddik2013disorder_vsr}. 

The state-of-the-art visual speech recognition (VSR) system relies on about 90K hours of transcribed training data~\cite{av-conformer2}. 
However, annotating such large amounts of data for every language is simply infeasible which sparked large interest to learn from unlabeled data.
AV-HuBERT~\cite{AV-HuBERT} was the first self-supervised system to jointly learn speech representations from raw audio and video using masked-prediction. 
However, the training is not entirely end-to-end since the algorithm alternates between representation learning and creating targets using offline clustering.
More recently, RAVen \cite{RAVen} introduced an end-to-end algorithm similar to data2vec~\cite{baevski2022data2vec} which trains separate encoder models for audio and visual data.
However, separate encoders increase the number of model parameters, and their disjoint model design is also contradictory to the common understanding of the human perception system which is believed to fuse audio and vision early on~\cite{green1998early_brain}.
Morever, they do not push the limit of AVSR which tends to perform better than ASR~\cite{shi2022robust_avhubert, hsu2022u-hubert}. 

\begin{figure*}[!htp]
    \centering
    \includegraphics[height=5.5cm]{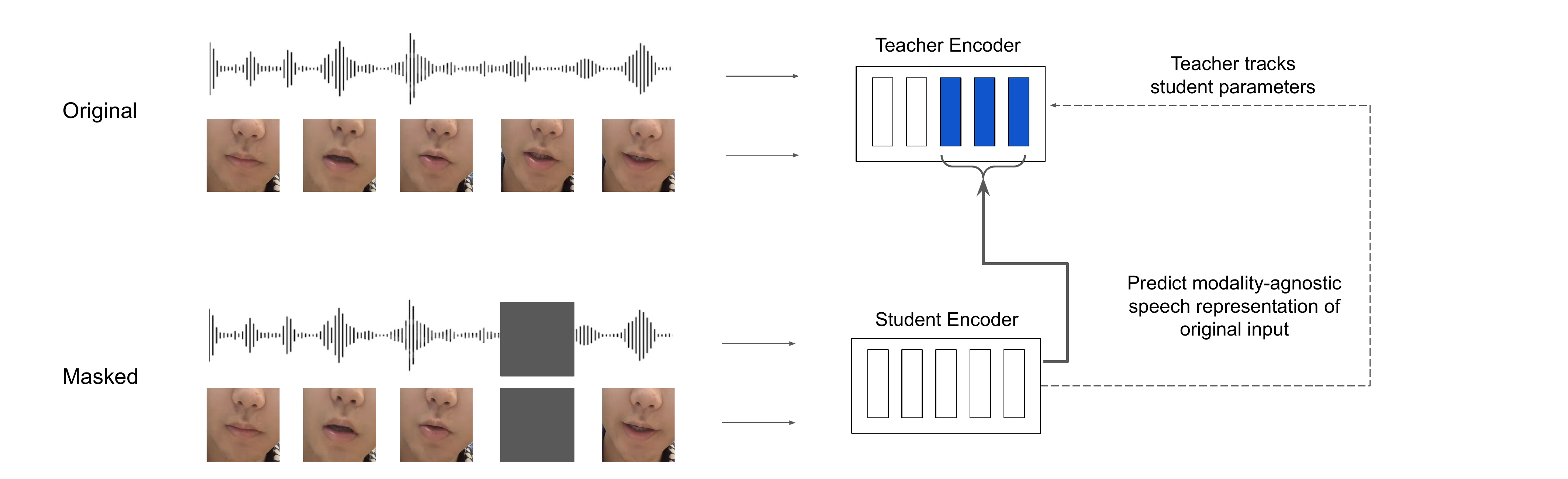}
    \caption{
    \nam{} jointly encodes both audio and visual data to build audio-visual representations. 
    The student model encodes a masked version of both audio and visual data and predicts a contextualized target representation created by a teacher model which is based on the unmasked version of the training sample. 
    Target representations encode both high-level and low-level features from multiple layers of the teacher model. 
    % In this example, the target representation is the average of the outputs of the last three layers.
    }
    \label{model_fig}
\end{figure*}

In this paper, we introduce \nam{} (Audio-Visual data2vec) to address these issues by extending data2vec \cite{baevski2022data2vec} from the unimodal case to learn joint audio-visual representations (\autoref{model_fig}).
\nam{} encodes masked audio-visual data and performs a masked prediction task of contextualized targets based on the unmasked input data.
Compared to prior work, training is fully end-to-end and there is a single encoder for both audio and vision that can be used to perform AVSR. 
Another difference to RAVen \cite{RAVen} is that target representations include features of varying granularity which is achieved by averaging the outputs of multiple layers instead of only predicting high-level features produced by the final layer.
This enables a learning task over both low-level and high-level features.
AV-data2vec unifies ASR, VSR and AVSR within a single framework and achieves state-of-the-art performance under all settings with the same amount of data/model size.

\section{Related Work}

\subsection{Self-supervised Speech Representation Learning.} 

There has been much recent research on self-supervised speech representation learning which includes approaches that reconstruct a corrupted or incomplete form of the input using auto-encoding~\cite{van2017vqvae}, auto-regressive based methods such as~\cite{chung2020apc, chung2020vqapc,ling2020dDecoar}, and masked prediction based methods~\cite{yue2021pMPC, liu2020npc}. 
% Another line of work maximizes the similarities of positive pairs of embeddings and minimizes the similarities of negative pairs of embeddings in the latent space~\cite{oord2018cpc,schneider2019wav2vec,baevski2019vqwav2vec,baevski2020wav2vec2}. 
There is also work on predicting the frame-wise targets outside of the model computational graph~\cite{baevski2019discreteBERT, hsu2021hubert, chen2022wavlm}. 
Related to the current paper is \cite{baevski2022data2vec, baevski2022data2vec2} who directly regress contextualized targets created by a teacher model. 
% Contextualization can capture the information of the entire sequence in the latent target representation. 

% \paragraph{Self-supervised Multi-modal Training.}
% A lot of recent attention has been devoted to self-supervised learning over multiple modalities such as vision (image and video), text, and speech. 
% Multi-modal training enables to ground each modality in another modality through a cross-modal learning task. 
% This line of work includes image-text pretraining~\cite{li2019visualbert, wang2021simvlm, radford2021clip,alayrac2022flamingo,dou2022meter,zeng2021xvlm,yang2022TCL,li2022flip}.
% Many explorations have also been made in video-text pretraining such as~\cite{sun2019videobert, luo2020univl, xu2021videoclip, luo2021clip4clip, fang2021clip2video}. 
% There are also numerous investigations taking place in the realm of speech tasks such as~\cite{bapna2021slam, ao2021speecht5, zhang2022speechut, chen2022maestro} in speech-text pretraining, \cite{chan2022avbert, AV-HuBERT,shi2022robust_avhubert,hsu2022u-hubert,RAVen} in speech-video pretraining, and \cite{akbari2021vatt, zellers2022merlot} speech-text-video pretraing. 

\subsection{Speech Recognition With Visual Cues.} 
Visual-oriented speech recognition involves the task of \textit{visual speech recognition} (VSR, also known as lip reading) and \textit{audio-visual speech recognition} (AVSR). 
Earlier work~\cite{afouras2018adeep, d-vsr, shillingford2018large-vsr, vsr-2022, rnn-vsr, prajwal2022sub-vsr, av-conformer1, av-conformer2} started to train with transcribed video/audio-video data in a supervised manner.
However, this required large amounts of labeled data of up to 90K hours~\cite{av-conformer2}.
There are some semi-supervised methods~\cite{cross-vsr, vsr-2022} which significantly reduce the amount of labeled data, however, the performance is still far lower. Most recent advances in self-supervised audio-visual learning \cite{AV-HuBERT, shi2022robust_avhubert, hsu2022u-hubert,RAVen} are not only more data-efficient but also achieve comparable or better speech recognition results.
AV-HuBERT \cite{AV-HuBERT} is the first method that jointly learns the modality-agnostic speech representation from raw audio and video. 
u-HuBERT \cite{hsu2022u-hubert} generalizes AV-HuBERT to utilize both multimodal and unimodal data that is richer in the wild during pretraining. 
% Additionally, it unifies fine-tuning with a modality-agnostic objective, achieving comparable or better results than modality-specific objectives. 
VATLM \cite{zhu2022vatlm} extends AV-HuBERT by adding auxiliary speech-text tasks which use additional out-of-domain text and speech data. One problem with these approaches is that multi-stage iterative training with offline clustered labels is not end-to-end.
RAVen \cite{RAVen} uses a student-teacher paradigm and is end-to-end, however, it uses separate encoders for each modality.
This is less parameter-efficient and very different to the human speech perception mechanism \cite{green1998early_brain}.
% Furthermore, AVSR \cite{shi2022robust_avhubert, hsu2022u-hubert}, which tends to outform ASR consistently, is not deployable in RAVen. 

\section{Method}

\subsection{Background: data2vec}\label{single-data2vec}

data2vec \cite{baevski2022data2vec,baevski2022data2vec2} is a self-supervised framework that learns the representations from \textit{contextualized} targets via masked prediction. 
Specifically, a student model encodes a masked version of the training example to predict a contextualized target representation encoded by a teacher model which is based on the unmasked version of the sample. 
The teacher model weights are an exponentially moving average (EMA) of the student model weights. The original data2vec framework is designed for single-modality training.

\subsection{Audio-Visual data2vec}\label{multi-modal data2vec}

We extend data2vec to multiple modalities and focus on speech and video inputs to create joint audio-visual representations (Fig.~\ref{model_fig}).
Similar to data2vec, \nam{} has a \textit{student encoder} and a \textit{teacher encoder}, however, instead of processing a single modality, encoders can represent both audio and visual data.
% The input to the student encoder is a masked version of both the audio and video data based on which the student network tries to predict contextualized target representations. These targets are created by a teacher encoder to which the unmasked version of the the same data is fed. 
Both the student and teacher networks are composed of an audio encoder $A$, a video encoder $V$, a audio-visual fusion module $F$ and a transformer encoder $T$.

\paragraph*{Audio Encoder.} 
Similar to \cite{AV-HuBERT}, we encode the audio signal as log filterbanks. 
We then adopt a dense layer as audio encoder $A$ that maps the $U$-frame log filterbank energy $X_{A}=[x_1,x_2,...,x_U]$ to acoustic features $M_{A}=[M_1,M_2,...,M_U]\in \mathbb{R}^{U\times D}$ of the same length: $M_A=A(X_{A})$.
The feature dimension $D$ of the audio encoder matches the input dimension of the transformer encoder. 
The audio feature $M_A$ is normalized per frame statistics for both pretraining and finetuning \cite{AV-HuBERT}.

\paragraph*{Video Encoder.} 
We use the same video encoder $V$ as AV-HuBERT which is a variant of ResNet-18 \cite{vsr-2022, AV-HuBERT,RAVen,zhu2022vatlm} that replaces the first 2D convolutional layer \cite{resnet} by a 3D convolutional layer with a kernel size [5, 7, 7] \cite{3dcnn}, followed by a batchnorm 3D layer \cite{ioffe2015batch}, a PRelu layer \cite{prelu} and a MaxPooling 3D layer with kernel size [1, 3, 3] and strides [1, 2, 2]. 
The visual features are then reshaped in order to be input to the subsequent 16-layer 2D convolutional layers \cite{resnet}. 
An adaptive average pooling 2D layer is applied in the end to output a 1D tensor for each frame. 
Given a $U$-frame raw video signal $X_{V}=[x_1,x_2,...,x_U]\in \mathbb{R}^{U\times C\times H\times W}$, the visual encoder $V$ maps $X_{V}$ to 1D visual features $M_{V}=[M_1,M_2,...,M_U]\in \mathbb{R}^{U\times D}$: $M_V=V(X_{V})$
Both the dimension $D$ and number of frames $T$ of visual encoder are the same as those of audio encoder. $C$, $W$, and $H$ denote channel, weight and height of each video frame. 

\paragraph*{Audio-Visual Fusion.}\label{audio-visual-fusion}

\nam{} accepts inputs that are either audio-only (a), video-only (v), or audio-video (av) for both student and teacher models. 
This leads to nine possible training tasks.\footnote{v$\rightarrow$a, av$\rightarrow$v, a$\rightarrow$a, v$\rightarrow$v, av$\rightarrow$v, a$\rightarrow$v, v$\rightarrow$av, av$\rightarrow$av, a$\rightarrow$av, where $\rightarrow$ denotes student-to-teacher prediction.} This compares to four learning tasks for RAVen \cite{RAVen} whose encoders can only encode a single modality each and which lacks the ability to jointly encode modalities. 
AV-HuBERT \cite{AV-HuBERT} can jointly encode modalities and uses \textit{modality dropout} to randomly select the type of input. 
In initial experiments, we found it very beneficial to adjust the rate at which each input type is selected over time during training.  

In this work, we propose a new modality scheduler that coordinates the nine different training tasks. 
We define the following parameters: $p_{A}$, $p_{V}$ and $p_{AV}$, denoting the probability that audio/video/audio-video is selected as input modality respectively for either the student or the teacher.\footnote{In the actual implementation, either audio or video is selected conditioned on audio-video not being selected. More precisely: $p_{A}=p_{\overbar{AV}}\;p_{A|\overbar{AV}}$ and $p_{V}=p_{\overbar{AV}}\;p_{V|\overbar{AV}}$, where $p_{\overbar{AV}} = 1 - p_{AV}$}

% Early results (Sec.\ref{ablations}) showed that audio-only targets performed best in our setup. 
We designed a modality dropout scheduler \textit{for the student model} where the rate at which modalities are dropout change over the time. 
% Denote $p^s_{V}$, $p^s_{A}$, $p^s_{AV}$, $p^s_{A|\overbar{AV}}$,  $p^s_{V|\overbar{AV}}$ as starting values and $p^e_{V}$, $p^e_{A}$, $p^e_{AV}$, $p^s_{A|\overbar{AV}}$,  $p^s_{V|\overbar{AV}}$ as ending values.
The probabilities $p_{AV}$, $p_{V|\overbar{AV}}$ and $p_{A|\overbar{AV}}$ are annealed: 
given a starting and an ending value for a probability, we linearly anneal the probability over $M_{anneal}$ steps. This results in $p_A$ and $p_V$ to be quadratically annealed over $M_{anneal}$ steps. 
% Details about these annealing strategy are in Appendix.\ref{appendix-modality-dropout}. 

% We design the linear annealing strategy for $p_{AV}$, $p_{A|\overbar{AV}}$ and $p_{V|\overbar{AV}}$. These strategies are formally defined as $p^s_{AV} \xrightarrow{M_{anneal}} p^e_{AV}$, $p^s_{A|\overbar{AV}} \xrightarrow{M_{anneal}} p^e_{A|\overbar{AV}}$ and $p^s_{V|\overbar{AV}} \xrightarrow{M_{anneal}} p^e_{V|\overbar{AV}}$ respectively. For example,  $p^s_{AV} \xrightarrow{M_{anneal}} p^e_{AV}$ meaning $p_{AV}$ linearly goes from $p^s_{AV}$ to $p^e_{AV}$ over $M_{anneal}$ steps. It is thus obviously to observe that both $p_V$ and $p_A$ follow the quadratic annealing strategy, which is detailed in Appendix.\ref{appendix-modality-dropout}. 

The audio-visual fusion module is summarized in Eq.~\ref{fusion}. Note that there are two independent audio-visual fusion modules for both the student model and the teacher model. 
\begin{equation}\label{fusion}
M=
    \begin{cases}
      M_A+M_V &\text{with probability } p_{AV}\\
      M_A+\textbf{0} &\text{with probability } p_{A}\\
      M_V+\textbf{0} &\text{with probability } p_{V}\\
    \end{cases}       
\end{equation}
If the input for both student and teacher model is audio-only data, then this is the same as data2vec \cite{baevski2022data2vec} framework for audio (A-data2vec; Sec.\ref{av-vs-a}). See supplemental material for a better understanding as well as more details for modality scheduler.

\paragraph*{Masking.}\label{masking}
Following \cite{baevski2020wav2vec2, hsu2021hubert}, we apply span masking on fused audio-visual features $M=[M_{1},M_{2},...,M_{U}]\in \mathbb{R}^{U\times D}$. 
We randomly select $r\%$ timesteps as starting indices to mask spans of length $l$. Note that if $M=M_A+M_V$, the masking is synchronously applied at the same time step for both audio and video, as illustrated in Fig.\ref{model_fig}. 

\paragraph*{Transformer Encoder.} 
The transformer encoder $T$ takes the masked fused audio-visual features $\Tilde{M}$ as input (cf. Eq.~\ref{fusion}) and outputs the high-level speech representation $Z=T(\Tilde{M})=[z_{1},z_{2},...,z_{U}]\in \mathbb{R}^{U\times D}$.

% Note that data2vec~\cite{baevski2022data2vec} unifies self-supervised learning across modalities with respect to learning objectives only. We take the step further and also unify the transformer encoder across audio and visual features in addition to loss functions. 

\subsection{Pretraining Objective}

\paragraph*{Targets.}
Similar to \cite{baevski2022data2vec}, \nam{} predicts contextualized targets  encoding a time-step as well as information about the entire input.
Targets are extracted from the representations encoded by the teacher encoder that takes the unmasked features as input. 
Following \cite{baevski2022data2vec}, we use the output of the FFN prior to the last residual connection in each block as target representation which is denoted as $\Bar{Z}\in \mathbb{R}^{U\times D}$. 
We furthermore denote the target representation at the last $k$ layer as ${\Bar{Z}}^{(N-k+1)}\in \mathbb{R}^{U\times D}$, where $N$ is the total number of transformer blocks, and $k$ is the current block.
We then average these representations over the last $K$ blocks and apply instance normalization similar to \cite{baevski2022data2vec} to derive the targets $Y=\text{IN}(\Sigma_{k=1}^K {\Bar{Z}}^{(N-k+1)})$, where $\text{IN}$ denotes instance normalization. 

\paragraph*{Loss.} 
Denote outputs of Transformer encoder $Z=[z_{1},z_{2},...,z_{U}]\in \mathbb{R}^{U\times D}$ and contextualized targets $Y=[y_1,y_2,...,y_U]\in 
\mathbb{R}^{U\times D}$. We consider computing our loss for both masked time-steps and unmasked time-steps \cite{RAVen}, depending on the input modality.
Empirically we find that audio-only targets perform best (See supplemental materials) and in this setting we found it useful to predict audio targets when we have visual-only inputs even for unmasked time-steps. 
Whenever we have video as input, then we only predict targets for unmasked time-steps as the task is otherwise trivial.
Specifically, if $t$ is the frame index, $I$ the set of masked indices, $\alpha$ and $\beta$ are two weighting factors, then the loss is: 
\begin{equation}\label{pre-loss}
    L_{pretrain}=\alpha \sum\limits_{t\in I} ||z_t-y_t||_2^2+\beta \sum\limits_{t\notin I} ||z_t-y_t||_2^2 
\end{equation}

\paragraph*{Teacher Parameterization}
Given student encoder weights $\theta_{S}$, the teacher weights $\theta_{T}$ are an exponentially moving average (EMA) similar to \cite{baevski2022data2vec}:
$$
    \theta_{T}\leftarrow \tau \theta_{T}+(1-\tau )\theta_{S}
$$
where $\tau$ is a momentum parameter that is linearly increased over time $\tau^s \xrightarrow{\tau_{anneal}} \tau^e$, where $\tau^s$, $\tau^e$ and $\tau_{anneal}$ denote the initial value, the ending value of EMA decay, and the EMA decay annealing steps.

\subsection{Finetuning Objective}
After pretraining, we initialize the encoder of an attention-based sequence-to-sequence (S2S) architecture \cite{s2s} and finetune it on labeled data.
% We also vary the transformer decoder size to explore the sensitivity of speech recognition performance with respect to decoder size. More details are discussed in Sec. \ma{Fix reference.} 
We denote the text targets as $W=[W_1,W_2,...,W_S]$ for the current input representation $Z=[Z_1,Z_2,...,Z_T]$. 
We minimize a cross-entropy (CE) criterion: $L_{S2S}=-\Sigma_{t=1}^{S}\log(W_t|W_{<t}, Z)$

\section{Experimental Setup}

\subsection{Datasets and Preprocessing}

\paragraph*{LRS3}~\cite{afouras2018lrs3} is the largest publicly available labeled dataset for audio-visual speech recognition in English. 
It is split as follows:
\textit{pretrain} (403h), \textit{trainval} (30h) and \textit{test} (1h). We follow~\cite{AV-HuBERT} to randomly select about 1h of data from \textit{trainval} for validation. 

\paragraph*{Voxceleb2}~\cite{chung2018voxceleb2} is a multilingual audio-visual dataset for speaker recognition without transcriptions. 
The original corpus contains more than 2442 hours of videos. 
We use the English-only part selected by~\cite{AV-HuBERT} (1326 hours of videos).  

\paragraph*{Preprocessing.}
For audio feature extraction, we follow~\cite{AV-HuBERT} and extract the 26-dimensional log filterbank energy with a stride of 10 ms from raw audio waveform. The original video track has a resolution of 224$\times$224 with a frame rate of 25 fps. 
Following~\cite{AV-HuBERT}, we use dlib~\cite{king2009dlib} to extract 68 facial key points for each video clip. 
We then crop a 96$\times$96 region centered on the speakers mouth. 
During training, we randomly crop a 88$\times$88 region from the whole region and flip it horizontally with probability 0.5. Following~\cite{AV-HuBERT}, we only take grayscale images. 
During testing, we use the 88$\times$88 region centered on the mouth and no flipping is applied. 
The frame rate for both modality is 25 fps. 
As the original audio features have a frame rate of 100 fps, we stack them for 4 audio features.

\subsection{Setup and Implementation Details}

We consider two experimental setups in terms of amount of labeled data: 
\textit{low-resource} and \textit{high-resource}. 
We pretrain \nam{} with either LRS3 (433h) or English-only Voxceleb2 + LRS3 (1759h). 
In the low-resource setting, the model is finetuned on LRS3 \textit{trainval} (30h) only and in the high-resource setting, the model is finetuned on the entire LRS3 training data (433h). 
Our methods are implemented in fairseq \cite {ott2019fairseq}. 
\paragraph*{Hyper-parameters Tuning}
The performance of AV-data2vec is sensitive to hyper-parameters such as how many blocks to average for the target representations, settings for the modality scheduler, EMA scheduler as well as batch size and learning rate. 
\paragraph*{Pretraining.}
Following \cite{devlin2018bert, AV-HuBERT, zhu2022vatlm}, there are two options for transformer encoder: Base and Large. The number of blocks/embedding dimension/feed-forward dimension/attention heads in each transformer block are \seqsplit{12/768/3072/12} and 24/1024/4096/16 for Base and Large respectively. For masking, we set mask probability $r\%=50\%$ and span length $l=10$. For pretraining loss defined in Eq.\ref{pre-loss}, we set $\alpha=1$ and $\beta=0$ if the input modality is audio-only or audio-video, and if the input modality is video-only, we set $\alpha=1$ and $\beta=1$. For student modality scheduler, we set 
$p_{AV}:1 \xrightarrow{150k} 0.25$, $p_{V|\overbar{AV}}:1 \xrightarrow{150k} 1$, $p_{A|\overbar{AV}}:0 \xrightarrow{150k} 0$. 
% The visualization of the scheduler is given in Appendix.~\ref{Scheduler5}.

For teacher modality scheduler, we set the input as audio-only. We fixed these modality schedulers for all pretraining experiments. For BASE model with 433h pretraining, we set lr=5e-4, $\tau^s=0.999$, $\tau^e=0.99999$, $\tau_{anneal}=100k$. The batch size is 20s per GPU. We set total number of updates as 1000k and the model is trained on 64 V100 for 4-5 days. For Base model with 1759h pretraining, we use almost the same settings with the exception that we double the effective batch size and train it for 2000k updates (8-10 days). For Large model with 433h pretraining, we still use the same settings as Base model with the exception that we double the effective batch size and lr=2e-4. It takes around 6-7 days to finish. For Large model with 1759h pretraining, we use the same settings as Base model with 1759h pretraining with the exception that we set lr=2e-4. It takes around 10-12 days to finish training. 

\paragraph*{Finetuning.}
We consider two transformer decoders: Base and Large. The number of blocks/embedding dimension/feed-forward dimension/attention heads in each transformer block are \seqsplit{6/768/3072/4} and 9/1024/4096/8 for Base and Large respectively. 
We use subword \cite{kudo2018subword} for S2S targets. For ASR/VSR finetuning, the video or audio features are set as zero vectors respectively. For AVSR finetuning, both video and audio are taken as input and there is no modality dropout. For ASR finetuning, we use tri-stage learning rate scheduler and freeze the encoder for some steps \cite{AV-HuBERT}. The learning rate/total number of updates/warmup steps for 30h/433h are 1e-3/1e-3, 40k/60k, 10k/20k, 24k/48k respectively. Settings are the same for both Base and Large model. 

For VSR finetuning, we use cosine learning rate scheduler and freeze the encoder for some steps \cite{AV-HuBERT}. The learning rate/total number of updates/warmup steps for 30h/433h are 1e-3/1e-3, 40k/120k, 2k/20k, 24k/48k respectively. Settings are the same for both Base and Large model. For AVSR finetuning, we use tri-stage learning rate scheduler and freeze the encoder for some steps \cite{AV-HuBERT}. The learning rate/total number of updates/warmup steps for 30h/433h are 1e-3/1e-3, 40k/60k, 10k/20k, 24k/48k respectively. Settings are the same for both Base and Large model. Since the AVSR results are not reported in \cite{AV-HuBERT} and are partially reported in \cite{shi2022robust_avhubert}, we reproduced AV-HuBERT and report our own AVSR results for the remaining settings, shown in Table.\ref{low-resource} and Table.\ref{high-resource}. 

\paragraph*{Decoding.} We tune the beam width in \seqsplit{$\{5,10,25,50,100\}$} and report the best number. We do not apply LM for decoding. For VSR, ASR and AVSR tasks, the input mdoalities are video-only, audio-only and audio-video respectively in both finetuning and decoding.

\begin{table*}[!htbp] 
    \centering
    \setlength{\tabcolsep}{10pt}

    \caption{Low-labeled Data Results. We pretrain \nam{} Large/Base with 433h/1759h of unlabeled data, and finetune on 30h of labeled data. 
    The results of visual speech recognition (VSR), automatic speech recognition (ASR) and audio-visual speech recognition (AVSR) are shown. CE denotes cross-entropy, also applying to Table.~\ref{high-resource}. \nam{} achieves state-of-the-art results in all settings with same amount of data/model size. 
    }

    \begin{threeparttable}
    
     \resizebox{16cm}{!}{
    \begin{tabular}
    {l r r r c r r r} 
     \hline
     Methods&Unlabeled AV data&Labeled Data&Encoder Size &Criterion&VSR&ASR&AVSR\\
 \rowcolor{Gray} 
 \multicolumn{8}{c}{\textit{Self-supervised (Base Models)}}\\
 
     AV-HuBERT~\cite{AV-HuBERT}& 433h & 30h&103M&CE&51.8 & 4.9 & 4.7\tnote{2}\\
     RAVen~\cite{RAVen} & 433h & 30h&97M&CTC+CE&47.0 & 4.7 & -\\
     VATLM~\cite{zhu2022vatlm} & 433h\tnote{1} & 30h&103M&CE&48.0 & - & \textbf{3.6}\\
     \rowcolor{brightturquoise} AV-data2vec & 433h & 30h&103M&CE&\textbf{45.2} & \textbf{4.4} & 4.2\\
     \hline
     AV-HuBERT~\cite{AV-HuBERT,shi2022robust_avhubert}& 1759h & 30h&103M&CE&46.1 & 4.6 & 4.0\\
     RAVen~\cite{RAVen} & 1759h & 30h&97M&CTC+CE&40.2 & 3.8 & -\\
     VATLM~\cite{zhu2022vatlm} & 1759h\tnote{1} & 30h&103M&CE&42.6 & - & 3.4\\
     \rowcolor{brightturquoise} AV-data2vec & 1759h & 30h&103M&CE&\textbf{37.8} & \textbf{3.7} & \textbf{3.3}\\
     \hline
     \rowcolor{Gray} \multicolumn{8}{c}{\textit{Self-supervised (Large Models)}}\\
     AV-HuBERT~\cite{AV-HuBERT}& 433h & 30h&325M&CE&44.8 & 4.5 & 4.2\tnote{2}\\
     \rowcolor{brightturquoise} AV-data2vec & 433h & 30h&325M&CE&\textbf{40.5} & \textbf{3.7} & \textbf{3.4}\\
     \hline
     AV-HuBERT~\cite{AV-HuBERT,shi2022robust_avhubert}& 1759h & 30h&325M&CE&32.5 & 2.9 & 3.3\\
     RAVen~\cite{RAVen} & 1759h & 30h&671M&CTC+CE&33.1 & 2.6 & -\\
     VATLM~\cite{zhu2022vatlm} & 1759h\tnote{1} & 30h&325M&CE&31.6 & - & 2.7\\
     \rowcolor{brightturquoise} 
     AV-data2vec & 1759h & 30h&325M&CE&  \textbf{30.8}& \textbf{2.7} & \textbf{2.7} \\
     \hline
    \end{tabular}}

     \begin{tablenotes} [para]
       \item [1] \scriptsize VATLM uses additional 3846h audio, 452h audio-text and 600M text data 
       \item [2] \scriptsize We reproduced AV-HuBERT and report our AVSR results. 
     \end{tablenotes}
    \end{threeparttable}
    \label{low-resource}
\end{table*}

\section{Results}

\subsection{Low-labeled Data Setup}

We first consider a low-labeled data setup using 30h of data for finetuning whose results are shown in Table.~\ref{low-resource}.
For Base models, \nam{} consistently outperforms existing methods for both VSR and ASR. 
On AVSR, \nam{} achieves the best result except for the 433h pretraining setting, where \nam{} achieves 4.2 compared to 3.6 for VATLM. 
However, VATLM~\cite{zhu2022vatlm} uses additional audio and text data for their auxiliary pretraining tasks. 
\nam{} appears to benefit more from an increased amount of pretraining data (1759h vs. 433h) than other approaches. 

For Large models, \nam{} achieves the best results except for the 1759h setting, where \nam{} gets 2.7 while RAVen gets 2.6. 
We attribute this in part due to RAVen having about double the model size due to their two encoder architecture.
Overall, with the same amount of pretraining data, larger models result in better performance. 
However, the benefits of increased model capacity and more pretraining data begin to diminish as can be seen in the results of the largest setting (Large model, 1759h pretraining data).

\begin{table*}[ht] 
    \centering
    \setlength{\tabcolsep}{10pt}
    \caption{High-labeled Data Results. We pretrain Base/Large models with 433h/1759h of unlabeled data and finetune on 433h of labeled data. Results of supervised/semi-supervised work are also included. AV-data2vec achieves state-of-the-art results under most settings. 
    } 
    \vspace{2mm}
    \begin{threeparttable}
     \resizebox{16.5cm}{!}{
    \begin{tabular}{l r r c c c r r r} 
     \hline
     Methods & Unlabeled &Labeled &Backbone&Encoder  &Criterion& VSR & ASR & AVSR\\
     & AV data & Data & & Size \\
     \rowcolor{Gray} \multicolumn{9}{c}{\textit{Supervised}}\\
     Afouras et al. 2018 \cite{afouras2018adeep}& - & 1519h&Transformer&-& CE& 58.9 & 8.3&-\\
     Xu et al. 2020 \cite{d-vsr}& - & 590h&RNN&-& CE& 57.8 & 7.2&-\\
     Shillingford et al. 2018 \cite{shillingford2018large-vsr}& - & 3886h&RNN&-&CTC & 55.1 & -&-\\
     Ma et al. 2022 \cite{vsr-2022}&- & 813h&Conformer&-&CTC+CE&34.7& - & -\\
     Makino et al. 2019\cite{rnn-vsr}& - &31000 &RNN&-& Transducer& 33.6 & 4.8&4.5\\
     Prajwal et al. 2022 \cite{prajwal2022sub-vsr}& - & 2676h&Transformer&-&CE&30.7 & - & -\\
     Serdyuk et al. 2021 \cite{av-conformer1}& - & 90000h&Transformer&-&Transducer&25.9 & - & 2.3\\
     Serdyuk et al. 2022 \cite{av-conformer2}& - & 90000h&Conformer&-&Transducer&17.0 & - & 1.6\\
     \rowcolor{Gray} \multicolumn{9}{c}{\textit{Semi-Supervised}}\\
    Afouras et al. 2020 \cite{cross-vsr}& 344h & 433h&Jasper(CNN)&-&CTC+CE&59.8 &-&-\\
     Ma et al. 2022 \cite{vsr-2022}& 641h & 818h&Conformer&-&CTC+CE&31.5& - & -\\
     \rowcolor{Gray} \multicolumn{9}{c}{\textit{Self-supervised (Base Models)}}\\
     AV-HuBERT~\cite{AV-HuBERT}& 433h & 433h&Transformer&103M&CE&44.0 & 3.0 & 2.8\tnote{2}\\
     RAVen~\cite{RAVen} & 433h & 433h&Transformer&97M&CTC+CE&39.1 & 2.2 & -\\
      \rowcolor{brightturquoise} AV-data2vec & 433h & 433h&Transformer&103M&CE&\textbf{39.0} & \textbf{2.0} & \textbf{1.8}\\
     \hline
     AV-HuBERT~\cite{AV-HuBERT}& 1759h & 433h&Transformer&103M&CE&34.8 & 2.0 & 1.8\tnote{3}\\
     RAVen~\cite{RAVen} & 1759h & 433h&Transformer&97M&CTC+CE&33.1 & 1.9 & -\\
     VATLM~\cite{zhu2022vatlm} & 1759h\tnote{1} & 433h&Transformer&103M&CE&34.2 & - & 1.7\\
      \rowcolor{brightturquoise} AV-data2vec & 1759h & 433h&Transformer&103M&CE&\textbf{32.9} & \textbf{1.7} & \textbf{1.4}\\
     \hline
     \rowcolor{Gray} \multicolumn{9}{c}{\textit{Self-supervised (Large Models)}}\\
     AV-HuBERT~\cite{AV-HuBERT}& 433h & 433h&Transformer&325M&CE&41.6 & 2.7 & 2.5\tnote{2}\\
      \rowcolor{brightturquoise} AV-data2vec & 433h & 433h&Transformer&325M&CE&\textbf{37.4} & \textbf{1.9} & \textbf{1.7}\\
     \hline
     AV-HuBERT~\cite{AV-HuBERT,shi2022robust_avhubert}& 1759h & 433h&Transformer&325M&CE&28.6 & 1.3 & 1.4\\
     RAVen~\cite{RAVen} & 1759h & 433h&Transformer&671M&CTC+CE&28.2 & 1.4 & -\\
     VATLM~\cite{zhu2022vatlm} & 1759h\tnote{1} & 433h&Transformer&325M&CE&28.4 & - & \textbf{1.2}\\
     u-HuBERT~\cite{hsu2022u-hubert} & 1759h\tnote{1} & 433h&Transformer&325M&CE&\textbf{27.2} & 1.4 & \textbf{1.2}\\
     \rowcolor{brightturquoise} AV-data2vec & 1759h & 433h&Transformer&325M&CE&28.5 & \textbf{1.3}& \textbf{1.3}\\
     \hline
    \end{tabular}}

     \begin{tablenotes} 
       \item [1] \scriptsize VATLM uses additional 3846h audio, 452h audio-text and 600M text data, and u-HuBERT uses additional 452h audio data.
       \item [2] \scriptsize We reproduced AV-HuBERT to report corresponding AVSR results. 
     \end{tablenotes}
     \end{threeparttable}
    \label{high-resource}
\end{table*}

\begin{table*}[ht] 
    \centering
    \setlength{\tabcolsep}{10pt}
    \caption{High-labeled Data Results. We pretrain Base/Large models with 433h/1759h of unlabeled data and finetune on 433h of labeled data. Results of supervised/semi-supervised work are also included. AV-data2vec achieves state-of-the-art results under most settings. 
    } 
    \vspace{2mm}
    \begin{threeparttable}
     \resizebox{16.5cm}{!}{
    \begin{tabular}{l r r c c c r r r} 

     Methods & Unlabeled &Labeled &Backbone&Encoder  &Criterion& VSR & ASR & AVSR\\
     & AV data & Data & & Size \\
     \rowcolor{Gray} \multicolumn{9}{c}{\textit{Supervised}}\\
     Afouras et al. 2018 & - & 1519h&Transformer&-& CE& 58.9 & 8.3&-\\
     Xu et al. 2020 & - & 590h&RNN&-& CE& 57.8 & 7.2&-\\
     Shillingford et al. 2018& - & 3886h&RNN&-&CTC & 55.1 & -&-\\
     Ma et al. 2022 &- & 813h&Conformer&-&CTC+CE&34.7& - & -\\
     Makino et al. 2019& - &31000 &RNN&-& Transducer& 33.6 & 4.8&4.5\\
     Prajwal et al. 2022& - & 2676h&Transformer&-&CE&30.7 & - & -\\
     Serdyuk et al. 2021 & - & 90000h&Transformer&-&Transducer&25.9 & - & 2.3\\
     Serdyuk et al. 2022 & - & 90000h&Conformer&-&Transducer&17.0 & - & 1.6\\
     \rowcolor{Gray} \multicolumn{9}{c}{\textit{Semi-Supervised}}\\
    Afouras et al. 2020 & 344h & 433h&Jasper(CNN)&-&CTC+CE&59.8 &-&-\\
     Ma et al. 2022 & 641h & 818h&Conformer&-&CTC+CE&31.5& - & -\\
     \rowcolor{Gray} \multicolumn{9}{c}{\textit{Self-supervised (Base Models)}}\\
     AV-HuBERT& 433h & 433h&Transformer&103M&CE&44.0 & 3.0 & 2.8\tnote{2}\\
     RAVen & 433h & 433h&Transformer&97M&CTC+CE&39.1 & 2.2 & -\\
      \rowcolor{brightturquoise} AV-data2vec & 433h & 433h&Transformer&103M&CE&\textbf{39.0} & \textbf{2.0} & \textbf{1.8}\\
     \hline
     AV-HuBERT& 1759h & 433h&Transformer&103M&CE&34.8 & 2.0 & 1.8\tnote{3}\\
     RAVen & 1759h & 433h&Transformer&97M&CTC+CE&33.1 & 1.9 & -\\
     VATLM & 1759h\tnote{1} & 433h&Transformer&103M&CE&34.2 & - & 1.7\\
      \rowcolor{brightturquoise} AV-data2vec & 1759h & 433h&Transformer&103M&CE&\textbf{32.9} & \textbf{1.7} & \textbf{1.4}\\
     \hline
     \rowcolor{Gray} \multicolumn{9}{c}{\textit{Self-supervised (Large Models)}}\\
     AV-HuBERT& 433h & 433h&Transformer&325M&CE&41.6 & 2.7 & 2.5\tnote{2}\\
      \rowcolor{brightturquoise} AV-data2vec & 433h & 433h&Transformer&325M&CE&\textbf{37.4} & \textbf{1.9} & \textbf{1.7}\\
     \hline
     AV-HuBERT& 1759h & 433h&Transformer&325M&CE&28.6 & 1.3 & 1.4\\
     RAVen & 1759h & 433h&Transformer&671M&CTC+CE&28.2 & 1.4 & -\\
     VATLM & 1759h\tnote{1} & 433h&Transformer&325M&CE&28.4 & - & \textbf{1.2}\\
     u-HuBERT & 1759h\tnote{1} & 433h&Transformer&325M&CE&\textbf{27.2} & 1.4 & \textbf{1.2}\\
     \rowcolor{brightturquoise} AV-data2vec & 1759h & 433h&Transformer&325M&CE&28.5 & \textbf{1.3}& \textbf{1.3}\\

    \end{tabular}}

     \begin{tablenotes} 
       \item [1] \scriptsize VATLM uses additional 3846h audio, 452h audio-text and 600M text data, and u-HuBERT uses additional 452h audio data.
       \item [2] \scriptsize We reproduced AV-HuBERT to report corresponding AVSR results. 
     \end{tablenotes}
     \end{threeparttable}
    \label{high-resource}
\end{table*}

\subsection{High-labeled Data Setup}

Results of the high-labeled data setting (433h) are shown in Table.~\ref{high-resource}. 
\nam{} achieves state-of-the-art \seqsplit{VSR/ASR/AVSR} results except for the largest setting (Large model, 1759h pretraining data):
u-HuBERT~\cite{hsu2022u-hubert} achieves the best VSR performanace of 27.2, however, it uses an additional 452h of data for pretraining.
VATLM~\cite{zhu2022vatlm} and u-HuBERT achieve the best AVSR results, however, VATLM uses additional 3846h audio, 452h audio-text and 600M text data, which gives it an advantage. In summary, AV-data2vec still achieves best results with the same amount of data/model size.

\subsection{Comparison to RAVen}

Similar to \nam{}, RAVen~\cite{RAVen} also uses contextualized and continuous targets, however, it differs from AV-data2vec in several important aspects. 
RAVen does not create joint modality embeddings and is not able to perform AVSR. 
Also, RAVen has different encoders for audio and video. 
For the Base model, each of the two encoders is half the size of \nam{} but collectively they have similar size. 
For the Large model, each RAVen encoder is the same size as \nam{} and thus the total size of RAVen in the Large size is about double of \nam{}.
Next, the finetuning criterion for RAVen is joint CTC-Attention \cite{watanabe2017hybrid} while \nam{} adopts a sequence to sequence architecture inline with AV-HuBERT \cite{AV-HuBERT} and VATLM \cite{zhu2022vatlm}.
Finally, \nam{} empirically performs better as our results show.

\subsection{Joint-modality vs. Audio-only Pretraining} 
\label{av-vs-a}

\begin{figure} \centering
    \includegraphics[width=60mm]{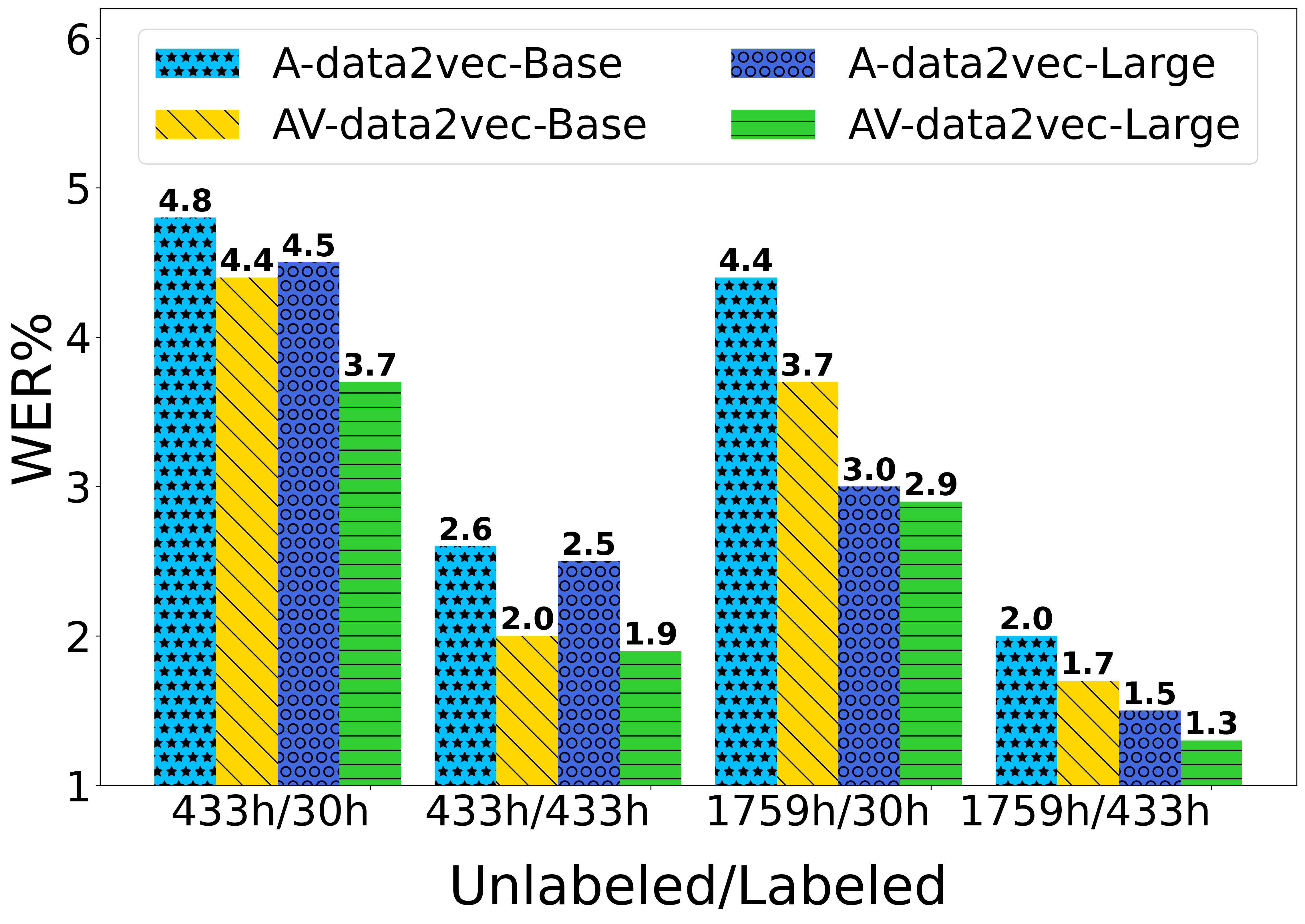}
    \caption{\nam{} performs better than audio-only training (A-data2vec) in all ASR settings.}
    \label{AV-training-A-training}
\end{figure}

Next, we compare joint audio-visual self-supervised learning to audio-only self-supervised learning.
To do so, we pretrain an audio-only version of our model (A-data2vec), by simply removing visual features before they are fed to the transformer encoder; we do not use modality dropout. 
We train A-data2vec for 600K updates for all settings and adopt the same finetuning/decoding configurations as \nam{}.
The ASR results (\autoref{AV-training-A-training}) show that joint audio-visual pretraining outperforms audio-only pretraining in almost all settings. 
In the largest high-resource setting (Large, 1759h unlabeled data, 433h labeled data), performance saturates and the difference to audio-only pretraining is very small.

\subsection{Ablation1: Top-K target averaging} \label{ablation-topk}
\begin{figure} [h]\centering
\includegraphics[width=70mm]{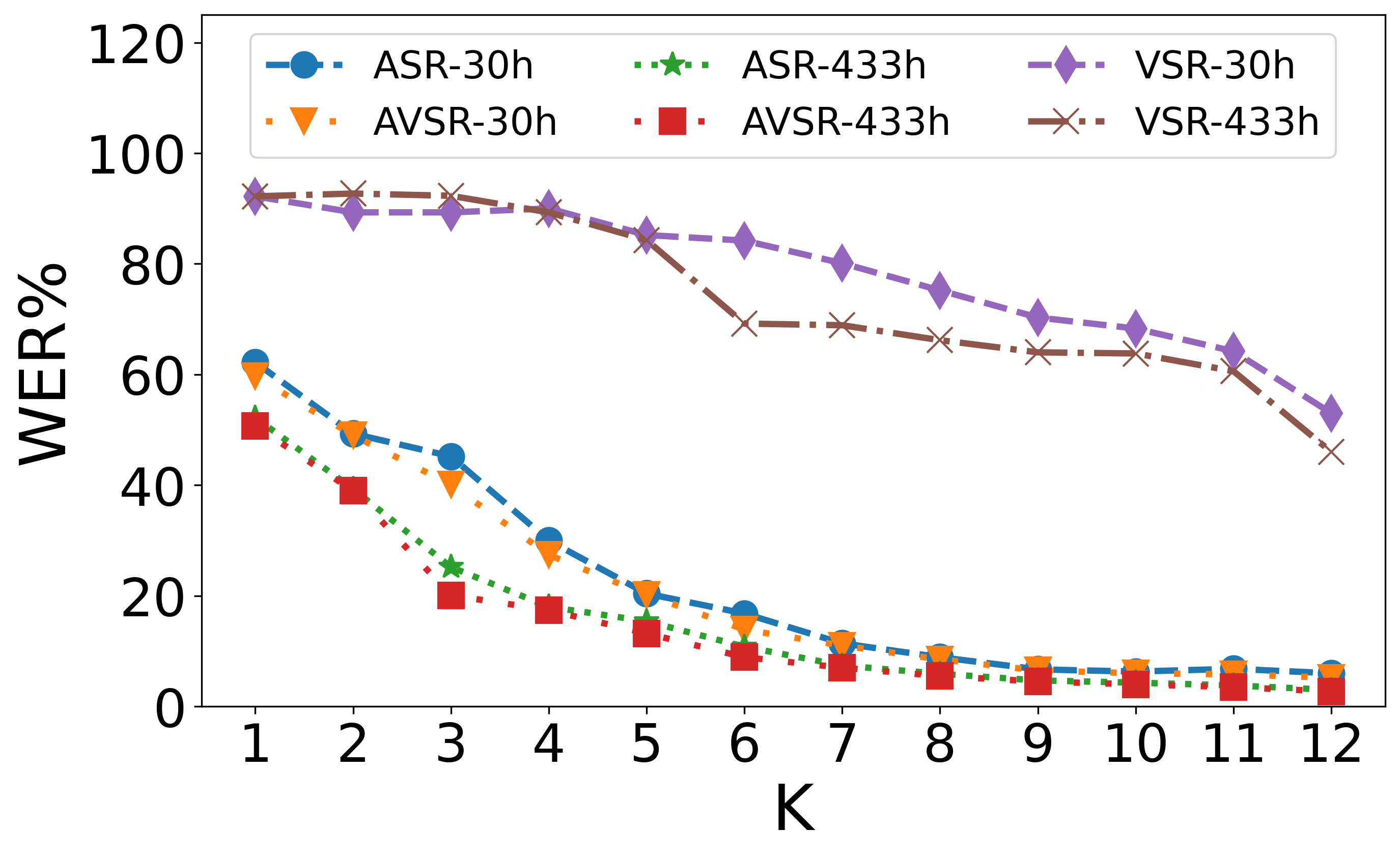}
\caption{Effect of averaging $K$ blocks to create contextualized target representations. 
More blocks improve performance because targets become richer due to including both high-level and low-level features.
Results are based on a Base model pretrained on 433h of unlabeled data and finetuned on 30h of labeled data.
}
\label{top-k-ablation}
\end{figure}

We first measure the impact of creating contextualized target representations based on multiple blocks ranging from the top block to the 12 blocks. 
For this experiment, we fix $p_{AV}=0.5$, $p_{A}=p_{V}=0.25$ for the student encoder which is the default schedule of \cite{AV-HuBERT}
we set $p_{A}=1, p_{AV}=p_{V}=0$ for the teacher encoder as video contains more ambiguous information as targets, as mentioned in~\cite{RAVen}. 
For EMA, we set $\tau^s=0.999$, $\tau^e=0.99999$ and $\tau_{anneal}=100k$. 
Fig.\ref{top-k-ablation} shows that averaging more blocks improves performance, inline with prior experiments for ASR, image recognition and natural language understanding~\cite{baevski2022data2vec}.
We therefore generally use $K=12$ for Base models and $K=24$ for Large models.

\subsection{Ablation2: Scaling}
For Large models and the largest unlabeled data setting (1759h), we investigate the effect of batch size and learning rates.
Table.\ref{scaling-dependent-ablation} shows the performance of a few settings we explored:
For 433h pretraining with Base model settings, increasding the batch size leads to plateauing performance.
However, when the amount of pretraining data is increased to 1759h, larger batch size still leads to better performance for all tasks.

For the Large model with 433h of unlabeled data, we found that smaller learning rates ($<$5e-4) improve performance; we find that 2e-4 gives the best performance.
When increasing the amount of pretraining data to 1759h, the largest batch size we considered (2560s) with learning rate 2e-4 performs very well.

\begin{table}[h]
    \centering
    \setlength{\tabcolsep}{5pt}
     \resizebox{8cm}{!}{
    \begin{tabular}{r r r r r r r r r r} 
     \hline
     \multicolumn{4}{c}{Configuration}&\multicolumn{3}{c}{30h Labeled Data} &\multicolumn{3}{c}{433h Labeled Data}\\
     \hline
     unlabeled&bsz &lr &model&VSR& ASR& AVSR &VSR& ASR& AVSR\\
     \hline

    433h&640s&5e-4&BASE& 48.7& 4.9& 4.7&40.6&2.2&2.0\\
    \rowcolor{brightturquoise}  433h&1280s&5e-4& BASE&45.2& 4.4& 4.2&39.0&2.0&1.8\\
     433h&2560s&5e-4& BASE&45.3& 4.5& 4.3&39.1&2.0&1.8\\

    1759h&640s&5e-4&BASE& 52.2& 4.9& 4.6&39.6&3.2&3.0\\
      1759h&1280s&5e-4& BASE&44.2& 4.2& 4.0&35.0&2.8&2.6\\
     \rowcolor{brightturquoise} 1759h&2560s&5e-4& BASE&37.8& 3.7& 3.3&32.9&1.7&1.4\\

    433h&1280s&5e-4& BASE&45.5& 4.3& 4.1&40.2&2.2&2.0\\ 
    \hline
    433h&1280s&3e-4& LARGE&43.7& 4.0& 3.8&39.8&2.0&1.9\\
    \rowcolor{brightturquoise}  433h&1280s&2e-4& LARGE&40.5& 3.7& 3.4&37.4&1.9&1.7\\
    433h&1280s&1e-4& LARGE&41.2& 3.9& 3.8&38.8&2.3&2.1\\

    \rowcolor{brightturquoise} 1759h&2560s&2e-4& LARGE&30.8& 2.7& 2.7&28.5&1.3&1.2\\

     \hline
    \end{tabular}}
    \caption{Ablation of batch size and learning rates for Base and Large models. bsz denotes batch size. Large models benefit more from smaller learning rates and larger amounts of unlabeled data benefits more from larger batch size. }
    \label{scaling-dependent-ablation}
\end{table}

\section{Conclusion and Limitations}

We proposed \nam{}, a self-supervised framework to jointly learn audio-visual speech representations based on contextualized targets. 
\nam{} adopts a shared modality-agnostic transformer encoder which takes as input both audio and video data, both of which are fused early on, similar to the human speech perception system. 
\nam{} unifies ASR, VSR and AVSR within a single framework and achieves state-of-the-art performance under all settings with the same amount of data/model parameters. Despite of this, there are still several limitations.  

Firstly, the current state-of-the-art self-supervised audio-visual speech recognition results are still inferior to supervised systems that rely on approximately 90K hours of labeled data \cite{av-conformer2}. Nevertheless, the self-supervised results for all of the current methods (AV-HuBERT \cite{AV-HuBERT}, VATLM \cite{zhu2022vatlm}, RAVen \cite{RAVen}, AV-data2vec) tend to reach saturation under high-resource and LARGE model settings. u-HuBERT \cite{hsu2022u-hubert} and VATLM attempt to use additional single-modality data to enhance performance, but the gain is limited.

Secondly, our results are sensitive to hyper-parameters, such as modality scheduler. The training process for both data2vec \cite{baevski2022data2vec} and AV-data2vec is not stable, which means that a good set of hyper-parameters can produce remarkable results. However, the optimal set of hyper-parameters may still be challenging to obtain. We believe that this high sensitivity is due to the fact that video data is much noisier than speech and contains less linguistic information. Furthermore, since our visual feature extractor, i.e., the ResNet-18, may not be capable of extracting sufficient useful information, the fused audio-visual feature may tend to be dominated by audio features. This sensitivity to the modality scheduler has also been observed in AV-HuBERT and RAVen. To address this issue, it would be beneficial to use a more powerful visual feature encoder such as Video Transformer that is adopted in VideoCLIP \cite{xu2021videoclip}. Additionally, implementing an information encoding monitoring method would provide better feedback for tuning the modality scheduler. It also worths to explore the audio-visual learning in the articulatory space~\cite{lian22bcsnmf, lian2023factor, inversion, wu23k_interspeech} to introduce vocal tract signal as additional signal to supervise the learning. 

\section{Acknowledgement}
We thank Bernie Huang for fruitful discussions around transformer block normalization schemes.
The purpose of this project is foremost to further the state of the art in audio-visual representation learning research.

%\vfill
%\pagebreak

% -------------------------------------------------------------------------
\bibliographystyle{IEEEbib}
\bibliography{strings,refs}

\end{document}